\documentclass[12pt,preprint]{aastex6} 
\usepackage{subfigure}
\usepackage{graphicx}
\usepackage{amsmath}

\slugcomment{Draft, revised 1, \today}

\shorttitle{New Suns in the Cosmos V}
\shortauthors{D. B. de Freitas et al.}

\begin{document}

\title{New Suns in the Cosmos V: Stellar rotation and multifractality in active \textit{Kepler} stars}


\author{D. B. de Freitas\altaffilmark{1}, M. M. F. Nepomuceno\altaffilmark{2}, L. D. Alves Rios\altaffilmark{1}, M. L. Das Chagas\altaffilmark{3} and J. R. De Medeiros\altaffilmark{4}}

\altaffiltext{1}{Departamento de F\'{\i}sica, Universidade Federal do Cear\'a, Caixa Postal 6030, Campus do Pici, 60455-900 Fortaleza, Cear\'a, Brazil}
\altaffiltext{2}{Universidade Federal Rural do Semi-\'Arido, Campus Central Costa e Silva, CEP 59600-900, Mossor\'o, RN, Brazil}
\altaffiltext{3}{Faculdade de F\'{\i}sica - Instituto de Ci\^encias Exatas, Universidade Federal do Sul e Sudeste do Par\'a, Marab\'a, PA 68505-080, Brazil}
\altaffiltext{4}{Departamento de F\'{\i}sica, Universidade Federal do Rio Grande do Norte, 59072-970 Natal, RN, Brazil}

\begin{abstract}
In the present study, high-precision time series photometry for the active \emph{Kepler} stars is described in the language of multifractals. We explore the potential of using the rescaled range analysis ($R/S$) and multifractal detrended moving average analysis (MFDMA) methods to characterize the multiscale structure of the observed time series from a sample of $\sim$40 000 active stars. Among these stars, 6486 have surface differential rotation measurement, whereas 1846 have no signature of differential rotation. As a result, the Hurst exponent ($H$) derived from both methods shows a strong correlation with the period derived from rotational modulation. In addition, the variability range $R_{var}$ reveals how this correlation follows a high activity ``line''. We also verify that the $H$-index is an able parameter for distinguishing the different signs of stellar rotation that can exist between the stars with and without differential rotation. In summary, the results indicate that the Hurst exponent is a promising index for estimating photometric magnetic activity.
\end{abstract}


\section{Introduction}
Stellar rotation is a fundamental parameter for investigating the magnetic fields in stellar interiors and spot dynamics on the stellar surface. Different indexes can be used to better understand stellar magnetic activity as a result from the interaction between rotation and convection.
Over more than 30 years, the spectroscopic index $S$-index developed by the Mount Wilson Observatory using $H-K$ flux variation has been used in studies on the correlation between magnetic activity and rotation measurements \citep{wilson1978,baliunas1995,mathur2014b}. This mission was a pioneer in validating models of stellar dynamos for stars with and without differential rotation traces. \cite{baliunas1995} showed that a large fraction of all main-sequence F, G, and K stars show cyclic variability and that this variability changes as a function of stellar age. In addition, these authors showed that the distribution of chromospheric activity depends on stellar mass as a function of the activity cycle variability. \cite{saar} used the Mount Wilson observations to explore the dependence of the amplitude of cyclic variability on different stellar parameters, such as $B-V$ color and effective temperature. The authors showed a steady increase in chromospheric Ca II HK emission \citep[related to $R_{HK}$-index from][]{noyes} with the $B-V$ index, where they observed that there was a decreasing of effective temperature in F and G stars until reaching a maximum in mid-K stars. Using photometric CoRoT mission data, \cite{garcia2010} defined a magnetic index by computing the standard deviation of the light curve using subseries of 30 days shifted every 15 days. 
Different methods can be used to measure rotation in spectroscopic or photometric contexts. For example, the rotational broadening of spectral line profiles is directly measurable by means of deconvolution techniques that are based on Fourier analysis \citep[e.g.,][]{Reiners2003} or Doppler imaging and Zeeman--Doppler imaging \citep[e.g.,][]{Strassmeier2009}. Recently, \cite{lanza2014} showed that differential rotation can also be extracted from the autocorrelation light curve (hereafter time series) for stars with marked starspots that are associated with simple two-spot models. 

\textit{Kepler} ultra-high precision photometry of long and continuous observations provides an unprecedented dataset to study the behavior of the rotation and stellar variability for almost 200 000 stars \citep{boru}. This opens a new perspective in the study of surface rotation. From the entire \textit{Kepler} database of time series for more than 800 stars observed in 17 quarters, \cite{daschagas2016} identified 17 stars with the signature of differential rotation and sufficiently stable signals. The authors used a simple two-spot model together with a Bayesian information criterion for this sample in the search to measure the amplitude of surface differential rotation \citep{lanza2014}. \cite{reinhold} noted that the \textit{Kepler} data allow us to measure differential rotation. Those authors used a procedure based on the Lomb-Scargle periodogram in a pre-whitening approach, resulting in a large sample of stars with differential rotation signature. They investigated a wide sample of 40 661 active stars and found 24 124 rotation periods between 0.5 and 45 days. This sample is based on Quarter 3, which was chosen because it has fewer instrumental effects than those for earlier quarters. In addition, the authors also found a second period in 18 616 stars that characterized the differential rotation signature.

In recent works, \cite{defreitas2013,defreitas2016,defreitas2017} have shown that multifractality analysis is a powerful tool for estimating correlations between stellar and statistical parameters, among them rotation period vs. the Hurst exponent, based on the geometric properties of the multifractality spectrum. More specifically, a set of four multifractal indices that are extracted from geometric features of the singularity spectrum (see Section 3) are used to describe the fluctuations in the different scales. The authors also show that the long-range correlation due to the rotation period of stars is scaled by the Hurst exponent, in agreement with Skumanich's seminal relationship \citep{sku}. 

Our main source of inspiration is based on the fact that most of the astrophysical time series exhibit self-similarity, which is the signature of a fractal nature in the system. Recently, \cite{defranciscis2018} used the fractal/multifractal frameworks to study the variability in the light curves of $\delta$ Scuti stars. Other works have been published \citep[e.g.,][]{elia,Bewketu} in this context, showing the strong applicability of multifractal analysis in the different astrophysical scenarios. 

In general, multifractal analysis and its different methods and procedures \citep{Kantelhardt,gu2010,tang}, which were developed over more than 5 decades, are applied in the most varied fields of knowledge as inspired by \cite{hurst1951,mw1969a,mw1969b,mw1969c,feder1988}. In several areas such as medicine \citep{ivanov1999} and geophysics \citep{telesca2006,db2008,defreitas2013b}, multifractality has already been adopted as a determinant approach for analyzing the behaviors of time series with nonlinearity, nonstationarity and correlated noise, to cite just a few of the properties that this analysis is able to describe \citep{movahed,Norouzzadeha,sps2009,seuront,a2011}. More recently, \cite{droz} showed that the multifractal analysis of sunspot numbers is a crucial procedure for understanding the behavior of the magnetic field of the Sun. Those authors also mentioned that the multifractal spectrum of sunspots is anomalous and of unknown physical origin. However, there are several approaches to investigate the self-similarity/fractality in the time series, such as Autocorrelation Function (ACF), Spectral analysis, Rescaled-Range analysis ($R/S$) and fluctuation analyses such as the Detrended Fluctuation Analysis (DFA) method and Multifractal Detrended Fluctuation Analysis (MF-DFA) \citep{Kantelhardt}. In the present paper, we will characterize magnetic activity in a sample of active stars by calculating the Hurst exponent through the $R/S$ \citep{defreitas2013} and MFDMA \citep{gu2010} methods. Our aim is to understand the physical mechanisms that drive the stellar magnetic activity. In this context, we proposed a new magnetic index based on the Kepler photometry that allow us to investigate the source of magnetic activity due to the presence of starspots on the stellar surface linked to the rotational period of the star.

In the present paper, we analyze the multifractal nature of an unprecedented sample of $\sim$ 40 000 active stars extracted from \cite{reinhold} and \cite{reinhold2015} with well-defined rotation periods and ages. To do so, we use the MultiFractal Detrending Moving Average (MFDMA) algorithm and Rescaled-Range analysis ($R/S$), both already tested by \cite{defreitas2016,defreitas2017} for the \textit{Kepler} and CoRoT stars. 

Our paper is organized as follows. In Section 2, we describe the working sample and methodology used. The required steps for producing the $R/S$ and MFDMA methods are introduced in Section 3 in which we emphasize a set of four indexes that are extracted from the multifractal spectrum. In Section 4, we define the Hurst exponent as a new photometric magnetic index. Section 5 is dedicated to a detailed discussion of the results. In the last section, our final remarks are summarized. 

\begin{figure}
	\includegraphics[width=0.9\columnwidth]{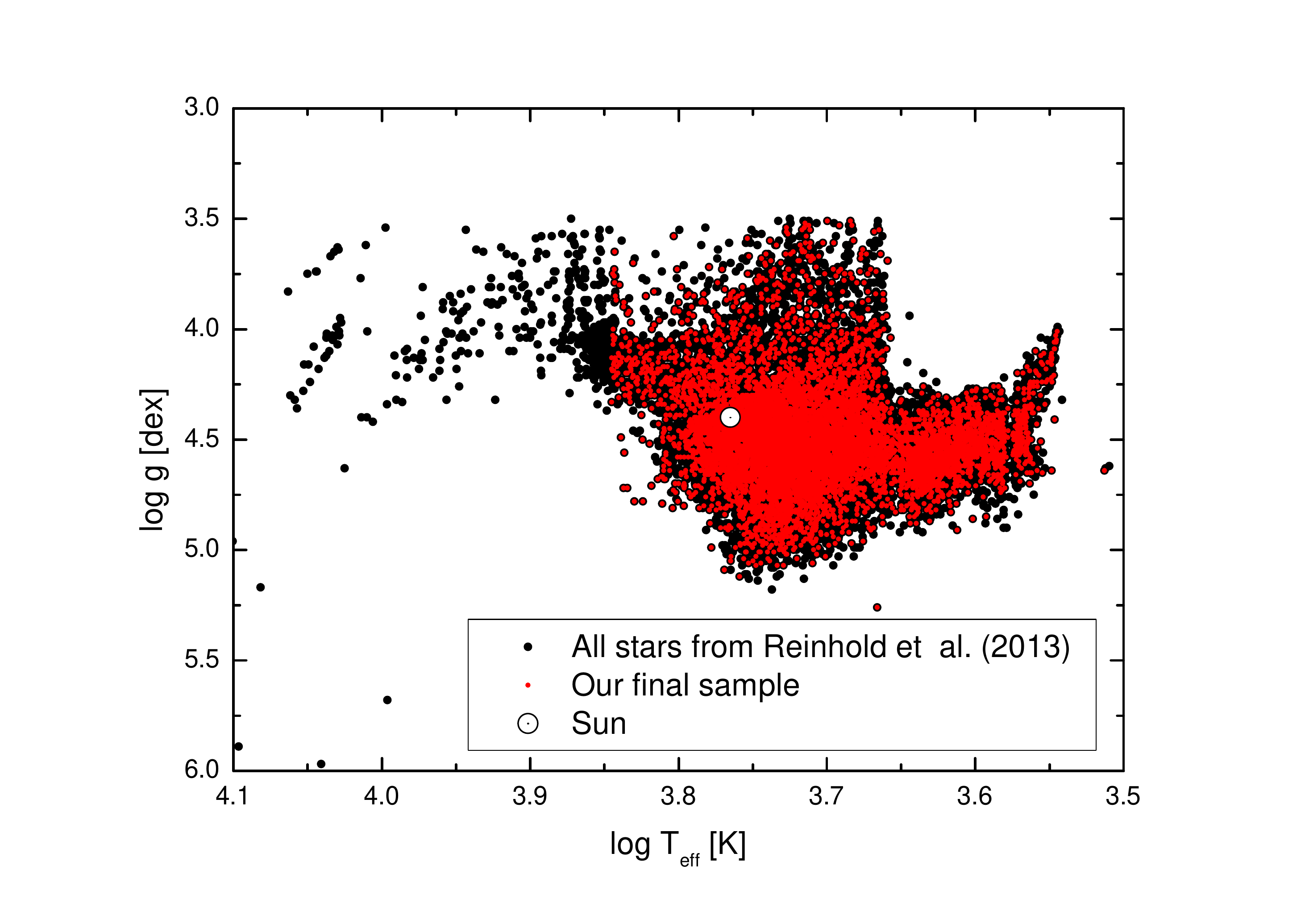}
	\caption{Effective temperature vs. the gravity of all the stars from \cite{reinhold}. }
	\label{HR}
\end{figure}

\begin{figure}
	\includegraphics[width=1.0\columnwidth]{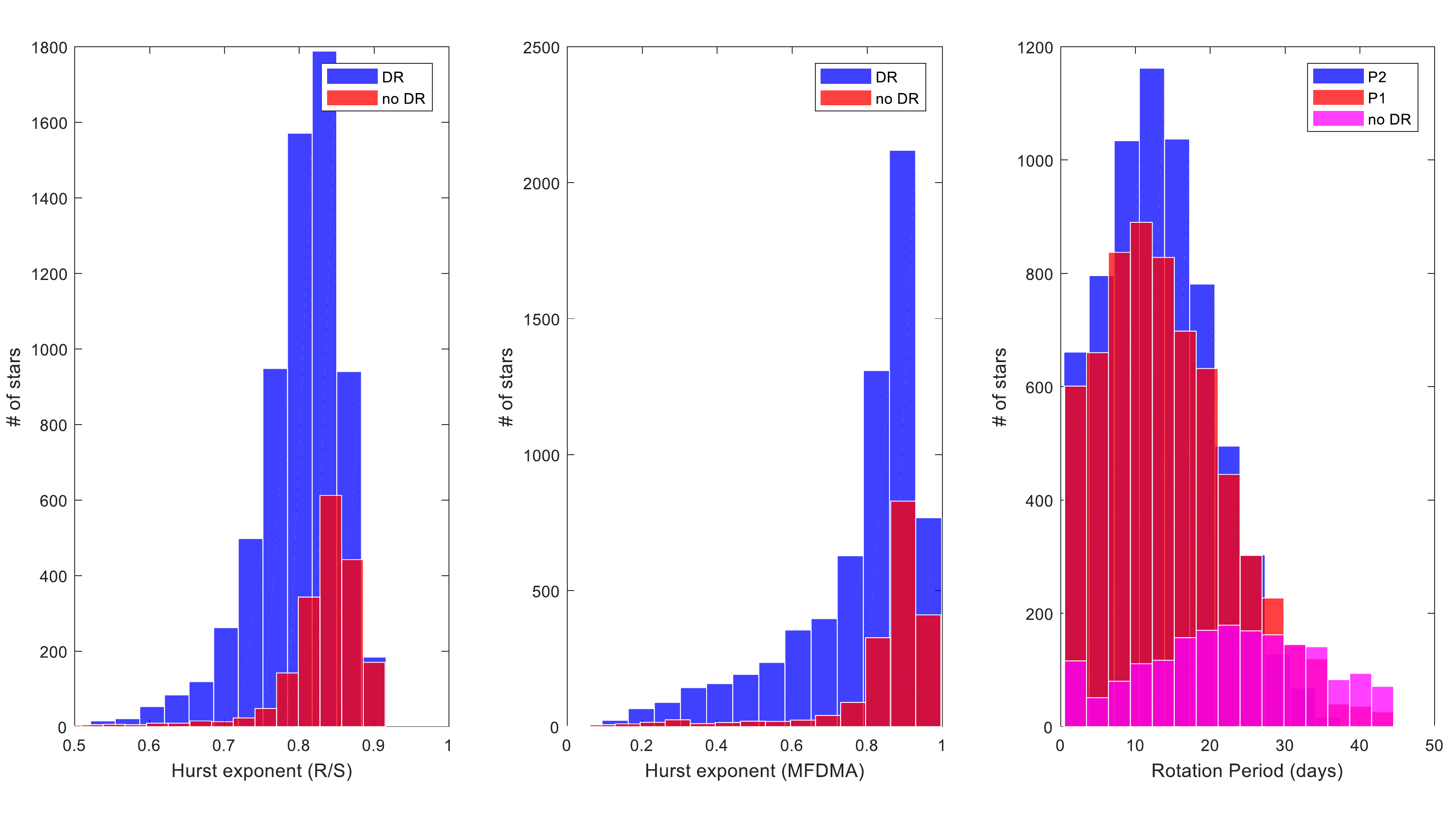}
	\caption{\textbf{Left panel}: The distribution of the Hurst exponent ($H$) via $R/S$ method for the 8332 Sun-like stars with differential rotation (DR) traces (blue histogram) identified by \cite{reinhold} and 1846 stars no DR (red histogram). \textit{Middle panel}: The same distribution for $H$ measured by MFDMA method. \textit{Right panel}: The distribution of the period for all stars of our sample.}
	\label{fig2}
\end{figure}


\section{Working sample and methodology}
The \emph{Kepler} mission performed 17 observational runs for $\sim$90 days, each of which was named by Quarters\footnote{\texttt{http://archive.stsci.edu/pub/kepler/lightcurves/tarfiles/}} and comprised long cadence (data \citep[data sampling every 29.4 min, see][]{Jenkins2010} and short cadence (sampling every 59 s) observations \citep{van,thompson}; detailed discussions of the public archive can be found in many \emph{Kepler} team publications, e.g., \cite{boru2009,boru}, \cite{batalha}, \cite{koch}, and \cite{basri2011}. A variety of pipelines have been used for processing the \textit{Kepler} time series. Initially, these data were processed by the Presearch Data Conditioning pipeline (PDC), which is not very careful when removing variability from the time series \citep{reinhold}. In the following, that pipeline was replaced by the PDC-MAP pipeline \citep{Jenkins2010b,stumpe2012,Smith2012}, and more recently, all Kepler data have been reprocessed by the PDC-msMAP (multiscale MAP) pipeline and implemented for long cadence data \citep{stumpe2014}. In addition, the PDC-msMAP pipeline reduced the chance that an astrophysical variability signature will be removed, consequently eliminating the systematic effects \citep{thompson}. As quoted by \cite{reinhold2015}, this new pipeline applies a 20-day high-pass filter, and as a consequence, it is not suitable for looking for stellar variability with a wide range of rotation periods because it diminishes stellar signals of slow rotators. For this study, we selected the calibrated time series processed by the PDC-msMAP pipeline \citep{garcia2014}. 

We applied the method developed by \cite{demedeiros2013} to remove outliers, a procedure that is able to identify exoplanet signatures and spurious points in the time series. However, we did not find marked differences between the indices calculated before and after this procedure. From this point on, the time series was considered to be fully treated, and fractal and multifractal analysis could be started.

Based on a working sample of 40 661 active stars adopted by \cite{reinhold} and \cite{reinhold2015} with rotation periods and ages that are well-determined, we constructed our time series using only Quarter 3 (Q3) long cadence data. From this sample of active stars, we selected 8 332 stars with main rotation periods shorter than 45 days. Our final sample is divided into 6 486 stars with surface differential rotation traces and 1 846 stars with no detected differential rotation signatures, defined by effective temperature $T_{\rm eff}$ shorter than 7000K. With this upper limit, we eliminate the periods that are most probably highly contaminated by pulsators. In addition, our sample of active stars was selected using the values of the variability range $R_{var}$ higher than 0.003 and shorter than 15$\%$ \citep{reinhold}. The detail procedure concerning $R_{var}$ can be found in \cite{reinhold}. All of the active stars occupy the dwarf regime with $\log g>3.5$. Figure \ref{HR} shows effective temperature vs. gravity of the \cite{reinhold} sample (black dots) with the stars selected in our sample shown in red. In addition, the values for the rotational periods were estimated using an auto-correlation function and were taken from \cite{reinhold}, and the temperature and gravity were obtained from \cite{pen} (SDSS corrected temperature and KIC surface gravity). In addition, our final sample roughly covers stars in the range of magnitude $8\lesssim K_{p}\lesssim 16$.

Another important parameter in our analysis is the variability range $R_{var}$. In a statistical sense, this parameter can be considered as an activity indicator. There are several measurements for describing photometric variability, and $R_{var}$ is one of them. In the present study, we used only $R_{var}$ as a photometric variability measurement, computed by \cite{reinhold} and using the quarter Q3 \citep{basri2010,basri2011}.

\section{Analysis methods}

In this Section, we describe two methods -- a fractal and another multifractal -- for analysing the \textit{Kepler} time series. Many methods for estimating the strength of the long-term dependence in a time series are available \citep{beran}. This “strength” can be measured by a seminal parameter called the Hurst exponent or self-similarity parameter. The parameter $H$ was initially developed by Harold E. Hurst while working as a water engineer in Egypt \citep{hurst1951} and introduced to applied statistics by \cite{mw1969a}, and it arises naturally from the study of self-similar processes \citep{barunik}. We chose the $R/S$ method to be one of the better known methods due to its robustness and computational and mathematical simplicity. On the other hand, the chosen multifractal method has become one of the promising methods found in the literature, and further details on its statistical efficiency are shown below.

\subsection{Rescaled range ($R/S$) analysis}
The well-known rescaled range ($R/S$) method is a simple but a strong nonparametric analysis for fast fractal analysis \citep{tanna}. In their work, \cite{defreitas2013} used this method proposed by \cite{mw1969b} for obtaining the global Hurst exponent $H$ using the following procedure. In general, a signal can be characterized by Hurst exponent $H$ defined by following empirical law \citep{hurst1951,defreitas2013,tanna}
\begin{equation}
\label{hurst4}
\frac{R(\tau)}{S(\tau)}=c  \tau^{H},
\end{equation}
where $c$ is a finite constant independent of $\tau$. In equation above, $s$ is the time lenght of the segment of the signal $y(t)$ and $R(\tau)$ is called the ``range'' and is given by expression
\begin{equation}
\label{hurst2}
R(s)=\max_{1\leq t\leq \tau} [Y(t,\tau)]-\min_{1\leq t\leq \tau} [Y(t,\tau)],
\end{equation}
where $Y(t,\tau)$ is defined as
\begin{equation}
\label{hurst3}
Y(t,\tau)=\sum^{t}_{n=1}[y(n)-\langle y\rangle_{\tau}],
\end{equation}
and
\begin{equation}
\label{hurstzero}
\langle y\rangle_{\tau}=\frac{1}{\tau}\sum^{\tau}_{t=1}y(t)
\end{equation}
where, $\langle y\rangle_{\tau}$ is the mean value of the signal over the time period $\tau$. $S(\tau)$ is the standard deviation of the signal and is defined by
\begin{equation}
\label{hurst}
S(\tau)=\left\{\frac{1}{\tau}\sum^{s}_{t=1}[y(t)-\langle y\rangle_{\tau}]^{2}\right\}^{\frac{1}{2}}.
\end{equation}

As argued by \cite{hurst1951}, the $R/S$ method is a powerful tool for detecting long-term memory and fractality of a time series when compared to more conventional approaches such as autocorrelation analysis. The Hurst exponent is obtained by the slope of the plot of $R/S$ versus the time span $s$ on a log-log plot \citep{defreitas2013}. The value of $H$ indicates whether a time series is random or whether successive increments in time series are not independent \citep{tanna}.

In particular, different values of $H$ imply fundamentally different variability behaviors on a time series. Values of $H$ equal to 0.5 show that a time series is an independent and identically distributed (i.d.d.) stochastic process, i.e., purely Brownian motion. For values between 0 and 0.5, a time series is anti-persistent, that is, the variability follows a mean reverting process. Finally, if $H$ is between 0.5 and 1, a time series is considered persistent with long-term memory. Broadly speaking, in a time series, if the dynamics that governs the variability is not known or if the signal is noisy, it is important to investigate the different sources of small and large fluctuations, as will be seen in Section 5.

\subsection{The multifractal analysis}
Several works, including \cite{MFDMAtemp}, \cite{Norouzzadeha} \cite{tanna} and \cite{defreitas2016,defreitas2017}, have applied multifractal analyses to time series as an effective statistical method to investigate the scaling properties of fluctuations. The most basic multifractal formalism \citep[see \textit{e.g.,}][]{MFstand} is based on the partition function, which fails when the series is non-stationary, meaning that it has a local trend or may not be normalized. Several methods have been created to improve the analysis of time series with these characteristics, including Wavelet Transform Modulus Maxima (WTMM) \citep[\textit{e.g.,}][]{WTMM1,WTMM2,WTMM3}, Multifractal Detrended Fluctuation Analysis (MF-DFA) \citep[\textit{e.g.,}][]{Kantelhardt} and Multifractal Detrended Moving Average (MFDMA) \citep[\textit{e.g.,}][]{gu2010}. Just as WTMM depends on the choice of wavelet function, MF-DFA depends on the choice of the degree of the local polynomial trend fit. The results for synthetic time series with compact support analysis have indicated that the WTMM and MF-DFA methods are equivalent, with MF-DFA offering a slight advantage for short series and negative values of {\it q} \citep{Kantelhardt}. A comparison of MF-DFA and MFDMA using synthetic series with known multifractal behavior can be found in \cite{gu2010}. MFDMA with a backward-moving average ($\theta = 0$) has been found to yield parameters with better alignments with the numerically calculated parameters, and this is the method employed in this work. Additionally, this value of $\theta$ has been demonstrated to achieve the best performance \citep{egh,defreitas2017,gu2010}.

Multifractal Detrended Moving Average is used here to calculate a set of multifractal fluctuation functions denoted by $F_{q}(s)$. According to the MFDMA procedure, we calculated the mean-square function $F^2_\nu(s)$ for a $\nu$ segment of size {\it s}. First, we have to divide the time series of length $N$ into series of the same size of $s$, when the number of windows is given by $N_{s}\equiv$int$(N/s)$. The fluctuations are calculated as sums of squares of local differences between the time series integrated over time $s$ and a time series detrended by removing the moving average function $\tilde{y}$. For $\nu \in\langle 1,N_s\rangle$ and $i \in\langle s,N\rangle$, we have

\begin{equation}\label{fluctuMS}
F^2_\nu(s)=\frac{1}{s}\sum_{i=1}^{s}[y(i)-\tilde{y}(i)]^2,
\end{equation}
We then calculated the $q_{th}$ order overall fluctuation function $F_q(n)$, which is given by

\begin{equation}
\label{eq4}
F_{q}(s)=\left\{\frac{1}{N_{s}}\sum^{N_{s}}_{\nu=1}F^{q}_{\nu}(s)\right\}^{\frac{1}{q}}, 
\end{equation}
for all $q\neq 0$, where the $q$th-order function is the statistical moment (\textit{e.g.}, for $q$=2, we have the variance), and for $q=0$,

\begin{equation}
\label{eq4b}
\ln\left[F_{0}(s)\right]=\frac{1}{N_{s}}\sum^{N_{s}}_{\nu=1}\ln [F_{\nu}(s)]. 
\end{equation}

For larger values of $s$, the fluctuation function follows a power-law given by

\begin{equation} \label{Fqxn}
F_q(s) \sim s^{h(q)},
\end{equation}
The generalized Hurst exponent \citep{hurst1951} $h(q)$ is a function of the magnitude of the fluctuations. Values of $h(q)$ are interpreted in three regimes: $0<h<0.5$ indicates antipersistency of the time series, $h=0.5$ indicates that the time series is an uncorrelated noise, and $0.5<h<1$ indicates persistency of the time series. Two other ranges are interesting: $h=1.5$ denotes Brownian motion (integrated white noise), and $h\geq 2$ indicates black noise.

The generalized Hurst exponent is related to standard multifractal analysis parameters such as the Renyi scaling exponent, $\tau(q)$, which is given by
\begin{equation}\label{tau}
\tau(q)=qh(q)-1.
\end{equation}

Finally, the multifractal spectrum is obtained using the Legendre transform to $h(q)$, defined as 
\begin{equation}\label{falpha}
f(\alpha)=q\alpha-\tau(q),
\end{equation}
with
\begin{equation}\label{alpha}
\alpha=\frac{d\tau(q)}{dq}, \quad \alpha\in[\alpha_{min},\alpha_{max}].
\end{equation}
In addition, for a monofractal signal, $h$ is the same for all values of $q$. For a multifractal signal, $h(q)$ is a function of $q$, and the multifractal spectrum is parabolic (see \cite{defreitas2017}, Fig. 2). In particular, the Hurst index ($H$) was obtained from the multifractal spectrum through the second-order generalized Hurst exponent $h(q = 2)$.

We use the following model parameters to yield the multifractal spectrum, as recommended by \cite{gu2010}: $q\in[-5,5]$ with a step size of 0.2; the lower bound of segment size $s$, which is denoted as $s_{min}$ and set to 10; and the upper bound of segment size $s$, which is denoted as $s_{max}$ and is given by $N/10$.

We calculated all four multifractal descriptors that were extracted from the spectrum $f(\alpha)$, as proposed by \cite{defreitas2017}. We recommend referring to Figure 2 elaborated by these authors to better understand each multifractal index. In the present work, we decided to investigate only the behavior of the Hurst exponent extracted by the two methods proposed here. This decision was based on the previous analysis of the correlations between this index and the stellar parameters available in \cite{reinhold} and \cite{reinhold2015}, as well as the previous studies from \cite{defreitas2013,defreitas2016,defreitas2017}, which point out the Hurst exponent as a promising classifier of rotational modulation. In this analysis, only the Hurst exponent presented strong correlations with the period of rotation, $R_{var}$ range, ages and amplitude of the differential rotation. The other indexes showed a weak correlation with these stellar parameters, and therefore, we believe that the work will be more impactful if we focus our efforts on the Hurst exponent. Thus, the correlations found for the degree of asymmetry ($A$), the degree of multifractality ($\Delta \alpha$), and the singularity parameters $\Delta f_{L}$, $\Delta f_{R}$ and $C$ are not shown here. In particular, the degree of asymmetry, which also called the skewness in the shape of the $f(\alpha)$ spectrum, is expressed as the following ratio:
\begin{equation}
\label{eq8}
A=\frac{\alpha_{max}-\alpha_{0}}{\alpha_{0}-\alpha_{min}},
\end{equation}
where $\alpha_{0}$ is the value of $\alpha$ when \textbf{$f(\alpha)$} is maximal. The value of this index $A$ indicates one of three shapes: right-skewed ($A>1$), left-skewed ($0<A<1$) or symmetric ($A=1$). The left endpoint $\alpha_{min}$ and the right endpoint $\alpha_{max}$ represent the maximum and minimum fluctuations of the singularity exponent, respectively \citep[further details, we recommend to see Figure 2 from][]{defreitas2017}.

\section{Hurst exponent as a new photometric magnetic index}
In the fractal context, the Hurst exponent is obtained by eq. (\ref{hurst4}). Already, in the multifractal one, the Hurst exponent is defined by the second-order statistical moment (i.e., variance or standard deviation) of $h(q)$, which is denoted by $q=2$ \citep[cf.][]{hurst1951,hurst1965,ihlen}. For both explanations of the exponent $H$, the reader can query \cite{defreitas2013,defreitas2017}. 	

Our main interest is to stand up a magnetic index to measure the degree of magnetic activity of stars with different rotational profiles from slow to fast rotators. To this end, we define a new photometric magnetic index like the Mount Wilson $S$-index denoted by the Hurst exponent $H$. This exponent is derived by different methods. Here, we showed the procedure for two of them: $R/S$ and MFDMA. In general, the $H$-index is sensitive to stellar variability and, as mentioned by \cite{defreitas2013}, is strongly correlated to the rotation period by a simple relation \citep[see eq. 1 in][]{defreitas2013}. Figure \ref{fig2} shows the behavior of the distributions for periods P1 and P2, and Hurst exponent $H$ calculated by two methods. Here, P1 and P2 are defined as first and second rotation periods, respectively. In the next section, the behavior of these distributions will be analyzed using a powerful statistical test.

\begin{figure}
\includegraphics[width=0.9\columnwidth]{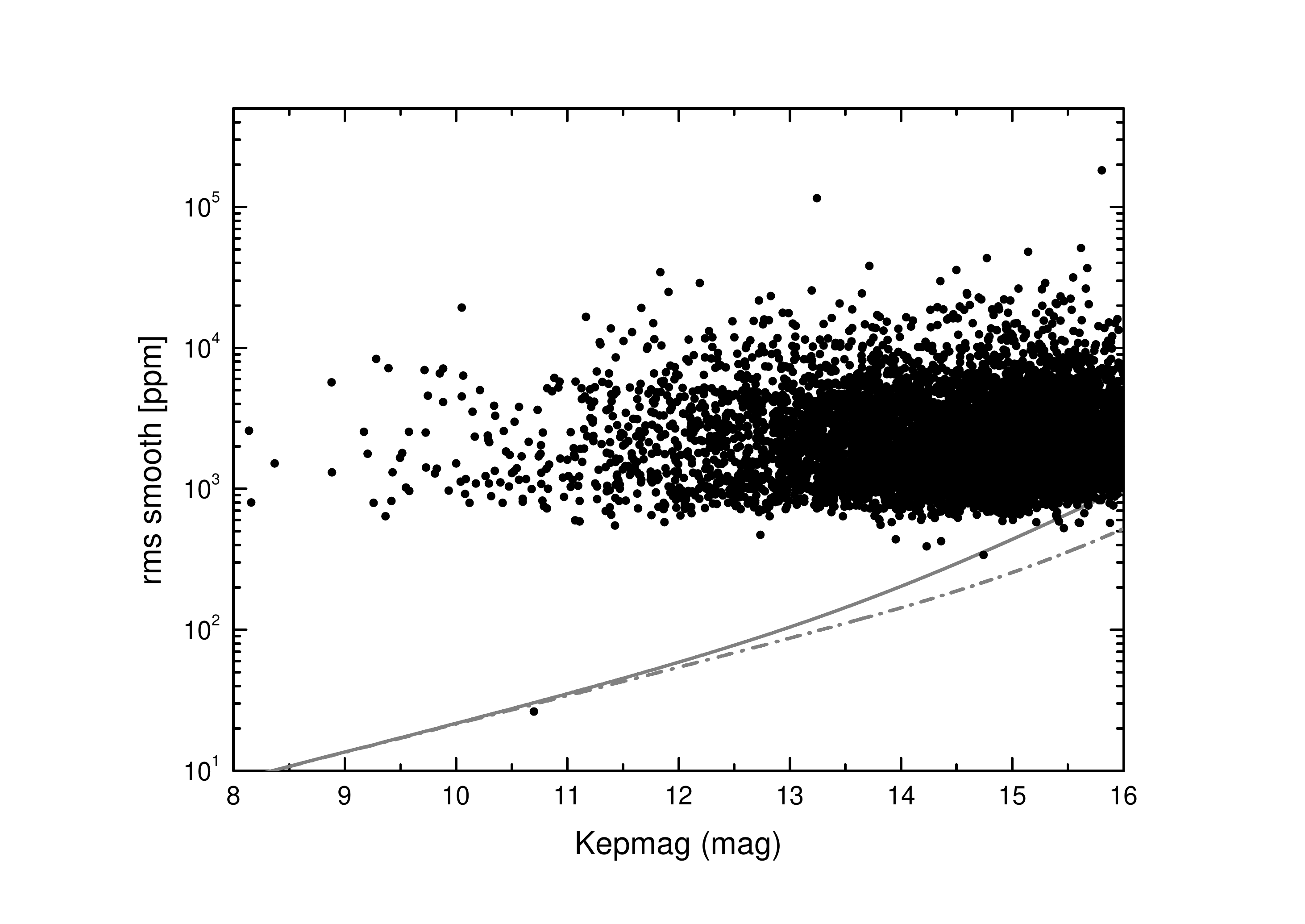}
\caption{The standard deviation of the smoothed time series for our final sample as a function of the \textit{Kepler} magnitude. The gray dash-dotted and solid lines indicate the lower and upper photon noise levels, respectively.}
\label{figPNoise}
\end{figure}

In general, the data can also be sensitive to photon noise. As mentioned by \cite{mathur2014a}, there are several ways to compute the influence of photon noise in data. We have followed the same procedure pointed out by these authors and use the methodology proposed by \cite{Jenkins2010b}. In addition, we calculated the minimum and maximum photon shot noise in the time series of the selected stars. Figure \ref{figPNoise} shows the standard deviation of the smoothed time series as a function of the \textit{Kepler} magnitude using the MATLAB function \texttt{smoothdata}\footnote{For more details, see https://www.mathworks.com/help/matlab/ref/smoothdata.html.} specifically adopted for noisy data. The gray dash-dotted and solid lines indicate the lower and upper photon noise levels, respectively, as defined by \cite{Jenkins2010b}. From figure \ref{figPNoise}, we conclude that there is no correlation between stellar variability and the apparent magnitudes of the stars. Moreover, all of the standard deviations of smoothed times series are above the estimated values for the contribution of photon noise.

\begin{figure}
\includegraphics[width=0.9\columnwidth]{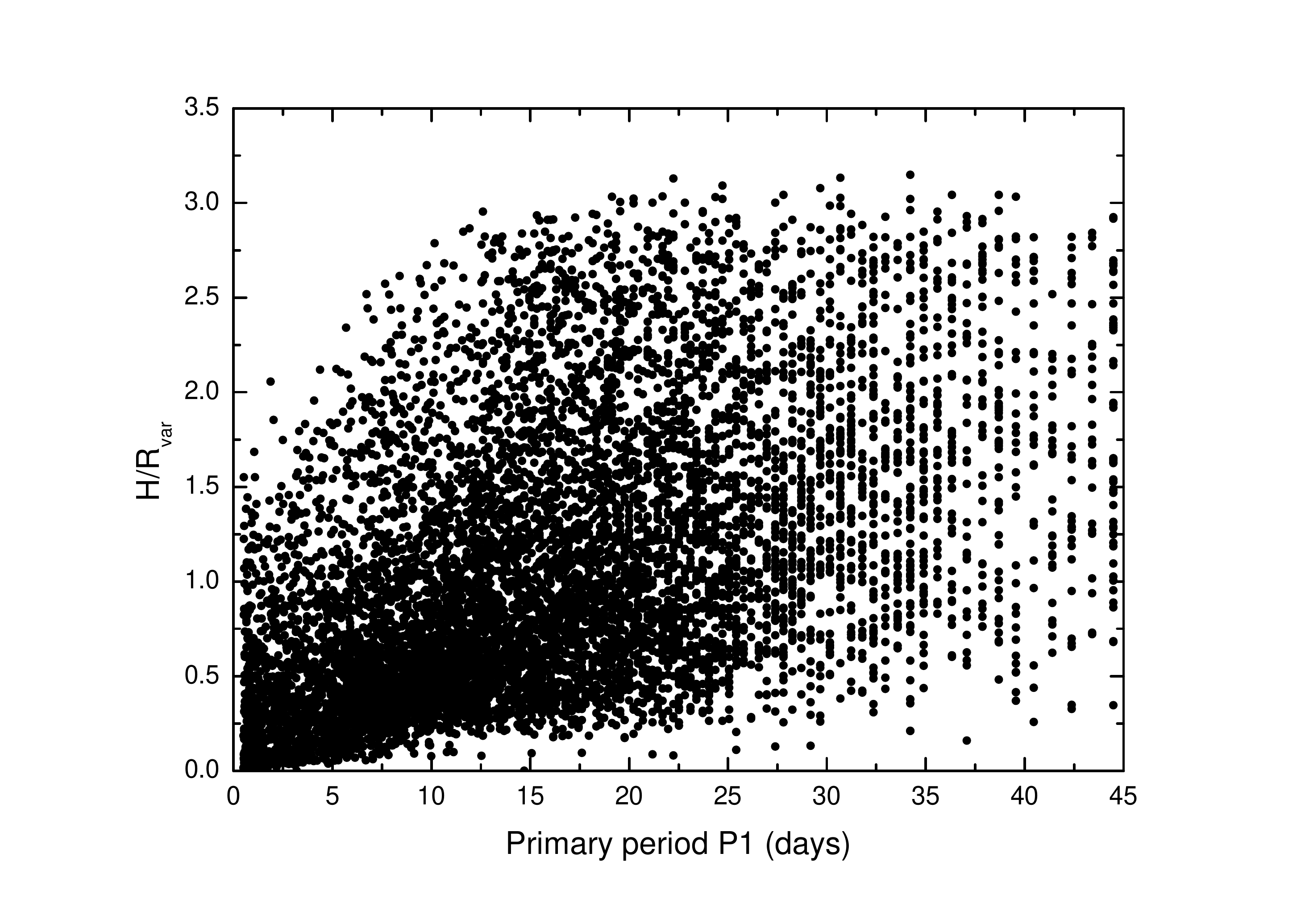}
\caption{Ratio between the multifractal $H$-index and the range $R_{var}$ as a function of the principal rotation period for stars with differential rotation traces observed by \textit{Kepler}.}
\label{figBias}
\end{figure}

\begin{figure}
\includegraphics[width=1.0\columnwidth]{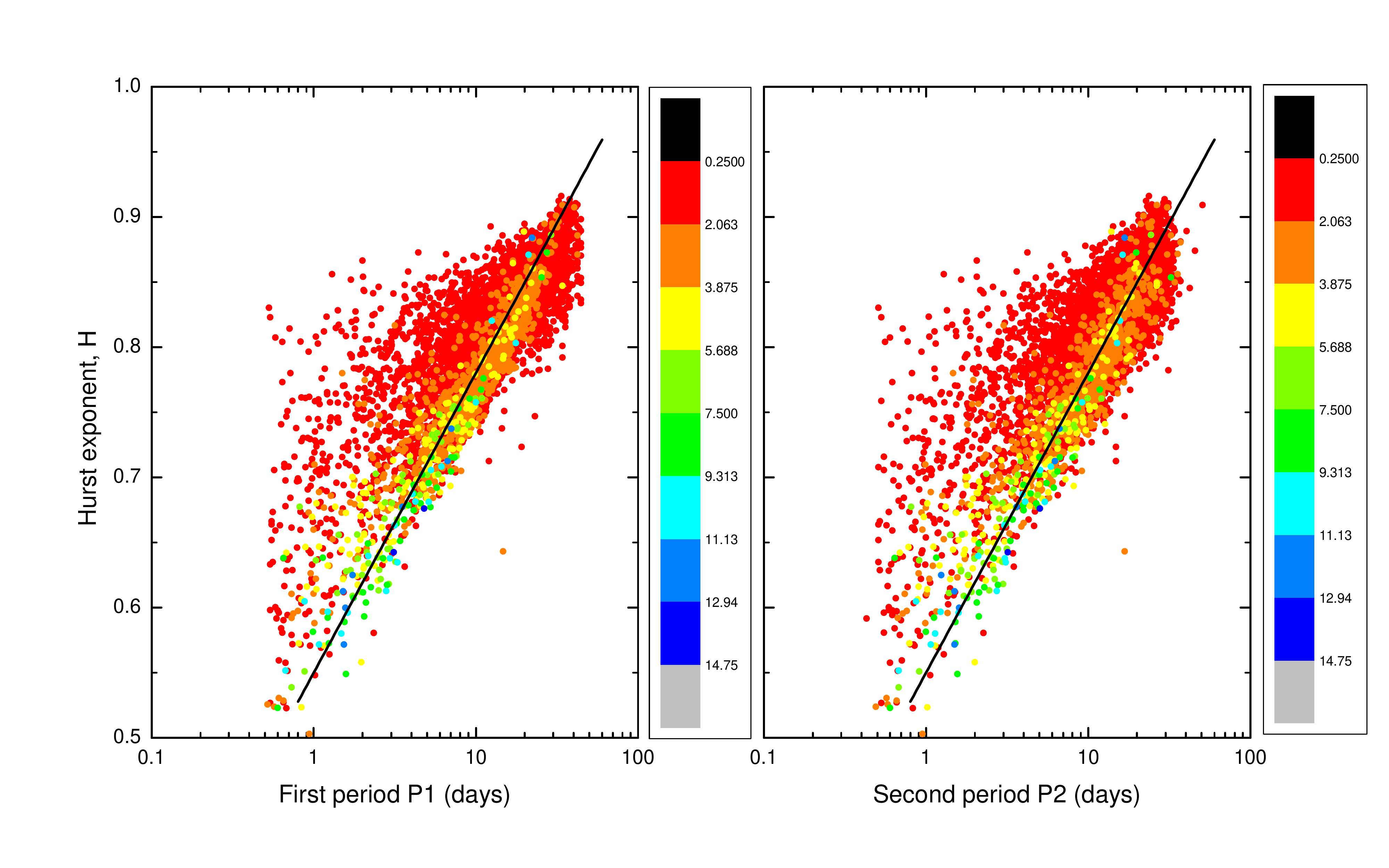}
\caption{Hurst exponent calculated by the fractal $R/S$ method vs. the first period P1 (left panel) and second period P2 (right panel) for all the stars with differential rotation traces. A clear high activity “line” (here denoted by black lines) is also presented in both panels of the figure according to the intensity of the variability range $R_{var}$. The color of the open circles is linked to the intensity of $R_{var}$, varying from 0.3 to 15$\%$ with an average value of 1.3$\%$ as shown in color scale. The black solid lines were extracted using eq. 1 from \cite{defreitas2013}.}
\label{figHFractal}
\end{figure}

\begin{figure}
\includegraphics[width=1.0\columnwidth]{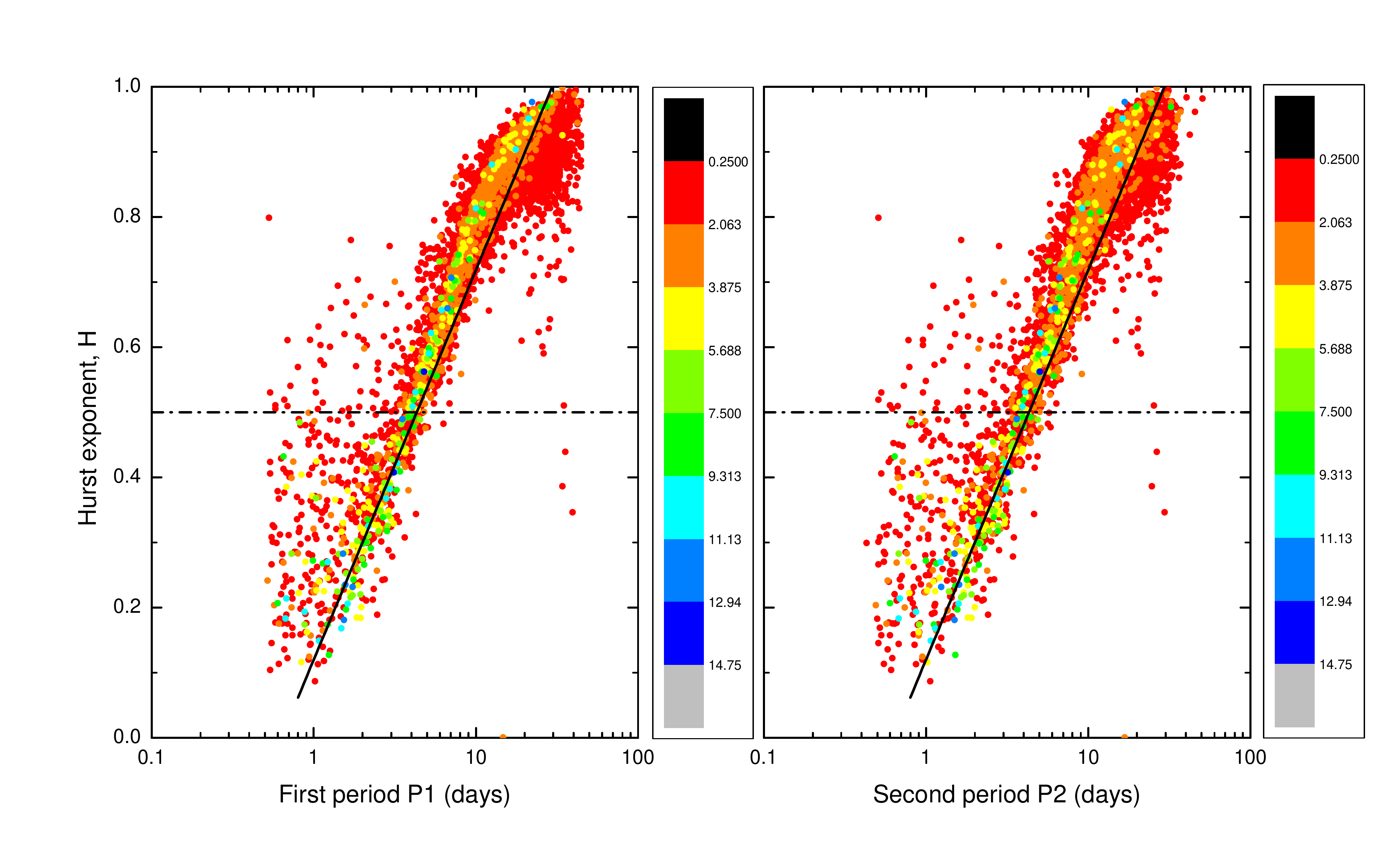}
\caption{Hurst exponent calculated by the multifractal MFDMA method vs. the first period P1 (left panel) and second period P2 (right panel) for all the stars with differential rotation traces. The figure also presents a clear high activity ``line'' (here denoted by black lines) in both panels according to the intensity of the variability range $R_{var}$. The color of the open circles is linked to the intensity of $R_{var}$, varying from 0.3 to 15$\%$ with an average value of 1.3$\%$ as shown in color scale. The black solid lines were extracted using eq. 1 of \cite{defreitas2013}. The horizontal dash-dotted line emphasizes the value of $H$ separating two persistence regimes.}
\label{figMFfractal}
\end{figure}

\begin{figure}

\includegraphics[width=1.0\columnwidth]{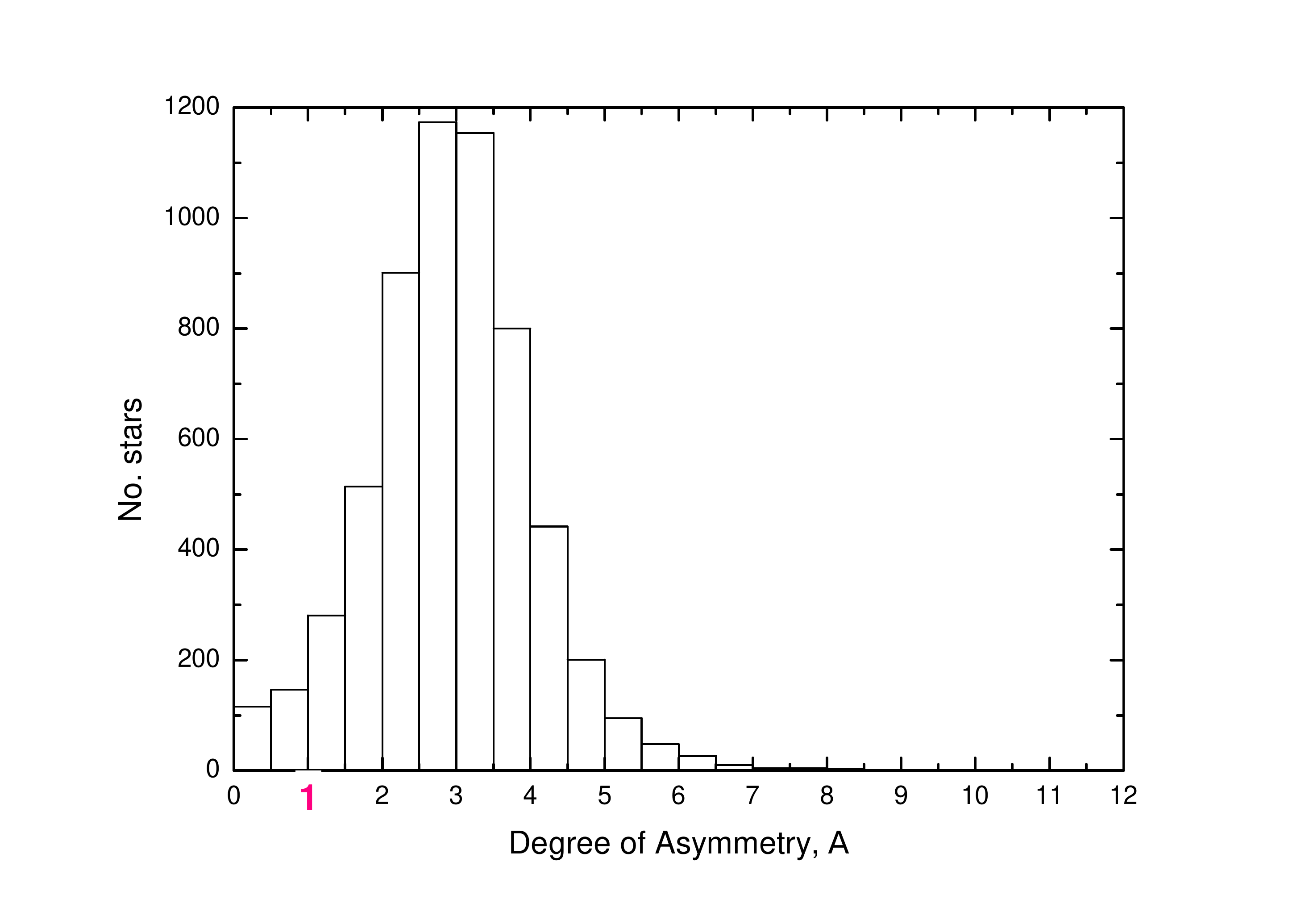}
\caption{Distribution of the degree of asymmetry for all stars. The red number is related to the value that separates the large-($A<1$) and short-($A>1$) magnitude fluctuations revealed by the long-right tail of the multifractal spectrum $f(\alpha)$}.
\label{figA}
\end{figure}

\begin{figure}
\includegraphics[width=0.9\columnwidth]{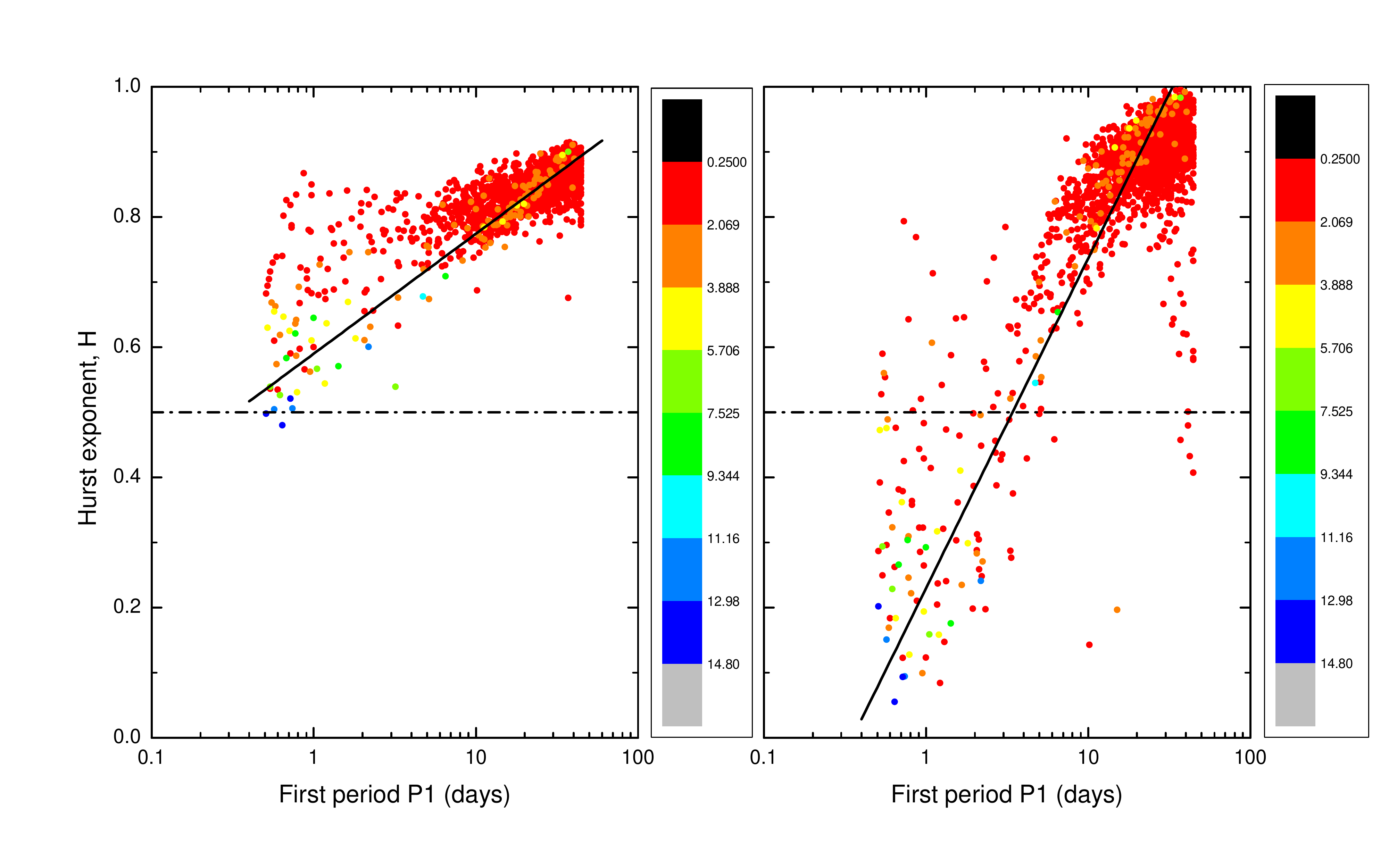}
\caption{The Hurst exponent calculated by the $R/S$ and MFDMA methods vs. the first period P1 for all the stars like rigid bodies. The figure also presents a soft activity ``line'' (here denoted by dark gray lines) according to the intensity of the variability range $R_{var}$. The color of the open circles are linked to the intensity of $R_{var}$, varying from 0.3 to 15$\%$ with an average value of 0.98$\%$ as shown in color scale. The gray lines were also extracted using eq. 1 from \cite{defreitas2013}, where $H=0.59+(0.08\pm 0.01)\ln(P_{rot})$ (left panel) and $H=0.23+(0.22\pm 0.02)\ln(P_{rot})$ (right panel).}
\label{figHr}
\end{figure}

\begin{figure}
\includegraphics[width=0.9\columnwidth]{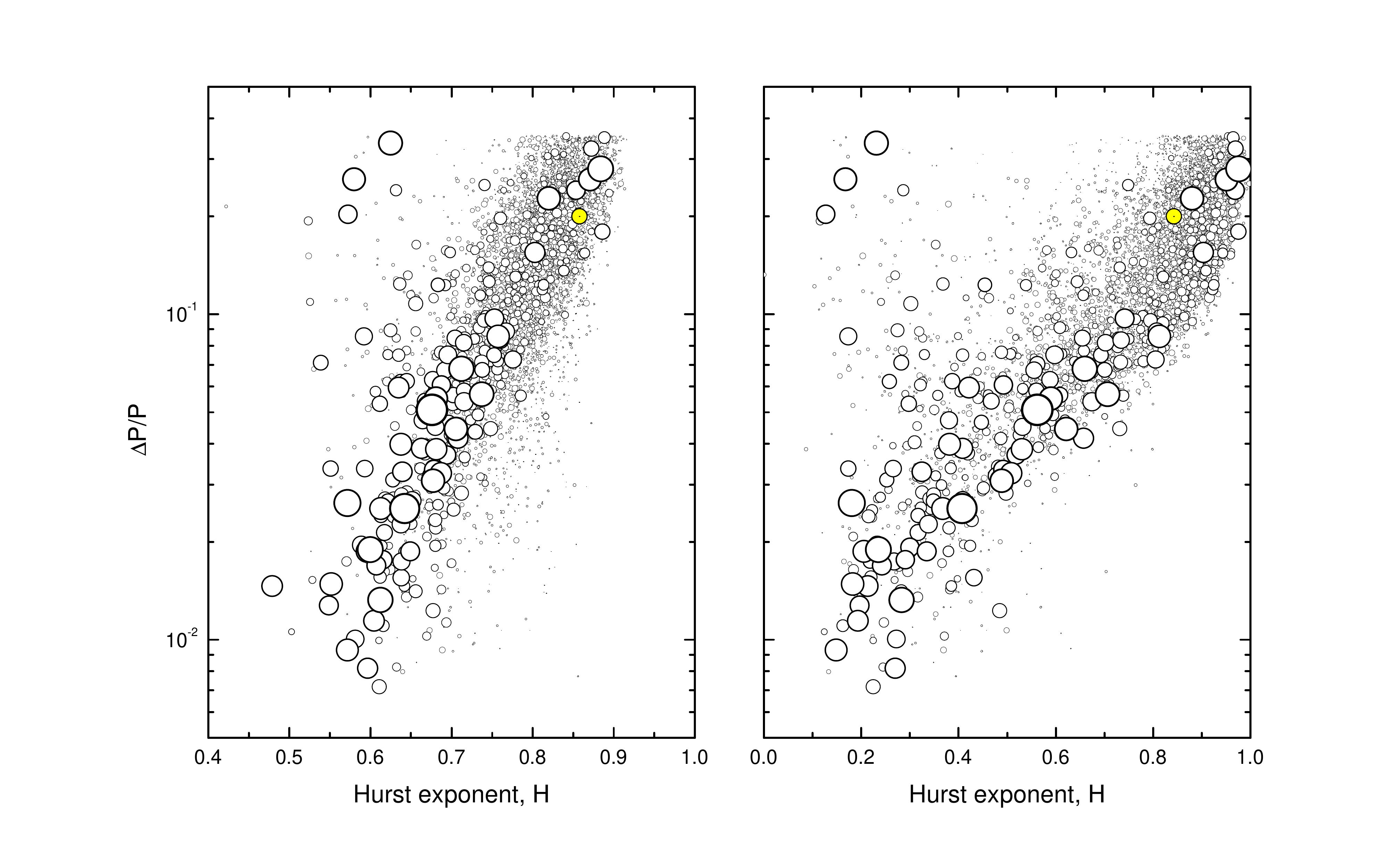}
\caption{The distribution of the relative amplitude $\Delta P/P$ versus the Hurst exponent $H$ for the stars with differential rotation traces identified in the present study. The left (right) panel was obtained using the $R/S$ (MFDMA) method for calculating the $H$-index. The solar values of $H$ (0.858) and $\Delta P/P$ (0.2) are denoted by the yellow symbol $\sun$. The circle size is proportional to the intensity of $R_{var}$ and is limited to range $0.3\leq R_{var} \leq 15$.}
\label{figHdeltaP}
\end{figure}

\section{Results and Discussion}

First of all, the $H$-index was calculated on segments of $\sim$ 90 days (1 quarter) of data. This limitation naturally could yield a bias towards slow rotators. Following the same procedure adopted by \cite{garcia2014}, we verify a possible presence of this bias using the $R_{var}$-index measured by \cite{reinhold}. Figure \ref{figBias} shows the ratio between the $H$-index and $R_{var}$ as a function of the rotation period. The figure shows clearly that no bias is introduced for stars with slower rotation periods. The same behavior occurs for $H$ measured by the fractal method (figure not shown).

Regardless of the method employed, the results are similar, showing a strong linear correlation between the $H$-index and rotation period in logscale (see Figures \ref{figHFractal} and \ref{figMFfractal}). Nevertheless, there is a clear distinction between the methods. The $R/S$ method is skewed to $H$ values higher than 0.5, whereas the MFDMA method extends over a wider spectrum of $H$. However, the behavior is similar, as can be seen in the mentioned figures. An explanation for this difference lies in the sensitivity to short- and long-term fluctuations in the analysis time series. In theory, the $R/S$ method is insensitive to the local fluctuations with large magnitudes and therefore favors the fluctuations with short magnitudes converging the values of $H$ to 0.5. In a wide study, \cite{barunik} mentioned that $R/S$ was shown to be biased for small $s$, and consequently, this behavior tends to overestimate the Hurst exponent. In contrast, the MFDMA method can distinguish these fluctuations efficiently \citep{defreitas2017}. Figure \ref{figA} makes this distinction clear, since the distribution of the degree of multifractal asymmetry points out that short magnitude fluctuations are dominant, and therefore, the signal of rotational modulation is stronger than background noise. Mostly, we found that the values of $A$ are greater than unity (see eq.~\ref{eq8}). This observation means that the fluctuations caused by larger magnitudes (background noise) are more likely to be monofractals than in the case of $A>1$.

Figures \ref{figHFractal} and \ref{figMFfractal} highlight an evident trail of high activity determined by the variability range $R_{var}$. The color of the open circles are linked to the intensity of $R_{var}$, varying from 0.3 to 15$\%$ as shown in color scale bar. Obviously behind the higher values of $R_{var}$ are the smaller ones located. However, our interest in this figure is to highlight the puzzling line that follows along the diagonal. In both methods, lower values of $R_{var}$ are clustered in the upper right corner. Another cluster of low $R_{var}$ values is concentrated in a triangular region located to the left of the high activity line. In general, the sample in the $H$-index versus rotation period semilog--plane is clearly divided into two domains, with the division along the values of $R_{var}$. The isolated stars with low $R_{var}$ in the triangular region of the figures tend to have high $H$, more rapid rotation, and therefore younger main-sequence ages (the age effect on the sample is analyzed below). The stars along the diagonal line vary on a wide spectrum of $H$, and therefore, the behavior between rotation and age follows a rotational decay curve as expected by \cite{sku}'s relationship. We checked all the correlations studied here using the Spearman and Pearson coefficients. Table 1 summarizes the values of these coefficients. On average, we see that the correlations are very strong, being above 0.7 for both statistical tests.

We also found that for values of $H$ below 0.7 (for the fractal method) and 0.5 (for the multifractal method), the population of stars with low $R_{var}$ values is drastically reduced. This is emphasized by the figures from \ref{figHFractal}, \ref{figMFfractal} and \ref{figHr} that highlight the stars that are in the background of the high activity line. As reported by \cite{reinhold2015}, the variability range from our sample more strongly decreases with age towards hotter stars. On average, this behavior is expected from the observation that young stars are more active than old ones. It is also expected that for the range of $H$, fluctuations with high magnitude are predominant and therefore accentuate the left tail of the multifractal spectrum \citep[see][for terminology]{defreitas2017}. As a result, we can underline that there is a cutoff at short periods ($<3$ days), where less active stars (low $R_{var}$) are not found on the diagonal line.

The figures do not show any meaningful difference in behavior between the first (P1) and second (P2) periods. Since the second period is not an alias or harmonic of period P1, the studied methods here can help us in confirming that the second period has a physical origin. This similar behavior reinforces that period P2 is not a statistical artifact and is therefore related to periodic variations caused by active regions located at certain latitudes. Given the present results, we are forced to think that if period P2 was not related to the dynamics of the active regions, figures \ref{figHFractal} and \ref{figMFfractal} would have a behavior outside this pattern. The gray lines present in the figures were obtained using eq. 1 from \cite{defreitas2013}, an equation like a linear fit of the form $H=a+b\ln(P_{rot})$, where $P_{rot}$ symbolizes both periods P1 and P2 given in days. Our best-fits point out that the period-$H$ relationship is given by

\begin{equation}
\label{dbf}
H=0.55+(0.10\pm 0.01)\ln(P_{rot}).
\end{equation}
In fact, the above equation is the same for both panels from Figure \ref{figHFractal}, and the equation below has the same profile for both correlations shown in Figure \ref{figMFfractal}:
\begin{equation}
\label{dbf2}
H=0.12+(0.26\pm 0.02)\ln(P_{rot}).
\end{equation}
As our goal is just to show the value of the slope to the highlighted trend, the intercepts were fixed.

The stars that spin like a rigid body were also analyzed. In this case, we do not find a correlation as clear as that revealed by stars with differential rotation (see Figure \ref{figHr}). However, the correlations and trends are close to the results found for stars with a differential rotation profile. Perhaps this discrepancy can be reduced if more of those stars are computed. In a future work, we will investigate this issue more deeply.

Here the analysis of the distributions of $H$ for stars with and without differential rotation is important. 
We use the two-sample Kolmogorov-Smirnov test (K-S test), where the null hypothesis assumes that the with and without differential rotation sample are from the same continuous distribution, whereas the alternative hypothesis indicates the opposite. In the present study, the K-S test is calculated using the MATLAB function \texttt{[h,p,k]=kstest2}\footnote{For more details, see https://www.mathworks.com/help/stats/kstest2.html}, where \texttt{h} can assume values 0 and 1, \texttt{p} is a probability used to reject or not the null hypothesis, and \texttt{k} is the maximum absolute difference between the maximum difference between the two cumulative distributions (cdf). If \texttt{h=1}, the test rejects the null hypothesis at the 1$\%$ significance level, and 0 otherwise. Likewise, if \texttt{p} is less than 1$\%$ significance level the null hypothesis can be rejected. We computed these values shown in Table 2. The values found in table indicate that the null hypothesis can be rejected at the 1$\%$ significance level. In addition, we find that the distributions of $H$ are different is quite high in all the two methods and, therefore, the two samples do not come from a common distribution.

\begin{deluxetable}{lcccccc}
	\tabletypesize{\scriptsize}
	\tablecaption{Spearman's (1st line) and Pearson's (2nd line) correlation coefficients ($r$). DRT means Differential Rotation Traces.}
	\tablewidth{0pt}
	\startdata
	& $P1$ & $P2$ & $R_{var}$ &  $\Delta P/P$\\
	& (days) & (days) & ($\%$)  & \\
	\hline 
	&		&	Stars WITH DRT	 & 	& \\
	\hline
	$H (R/S)$	&	0.83	&	0.79	&	-0.44	& 0.17 \\
	$H (R/S)$	&	0.82	&	0.80	&	-0.49	& 0.39 \\
	\hline
	$H (MFDMA)$	&	0.89	&	0.87	&	-0.19	& 0.70\\
	$H (MFDMA)$	&	0.92	&	0.92	&	-0.30	& 0.76\\
	\hline 
	&		&	Stars WITH NO DRT	 & 	& \\
	\hline
	$H (R/S)$	&	0.74	&	--	&	-0.21	&  --\\
	$H (R/S)$	&	0.80	&	--	&	-0.55	&  --\\
	\hline
	$H (MFDMA)$	&	0.61	&	--	&	0.04	&  --\\
	$H (MFDMA)$	&	0.83	&	--	&	-0.40	&  --\\
	\enddata
	\label{tab2}
\end{deluxetable}

\begin{table}
	\caption{Result of the Kolmogorov-Smirnov test giving the parameters \texttt{h}, \texttt{p} and \texttt{k} from the two cumulative distribution functions of $H$ of stars with and without differential rotation for each method.}
	\label{tab1}
	\begin{center}
		\begin{tabular}{lccccc}
			\hline
			Method & \texttt{h} & \texttt{p} & \texttt{k} \\
			\hline
			R/S & 1 & $<$0.01 & 0.30 \\
			MFDMA & 1 & $<$0.01 & 0.25	\\
			\hline
		\end{tabular}
	\end{center}
\end{table}

We also investigated the behavior of the relative amplitude $\Delta P/P$ as a function of the $H$-index for our sample stars with differential rotation traces determined by \cite{reinhold}, where $P$ is the spot rotation period computed by the mean values of the individual rotation periods P1 and P2; hence, $P = (P1 + P2)/2$ \citep{daschagas2016}. In contrast to the sample adopted by \cite{daschagas2016}, we analyze this correlation for a wide range of rotation periods, namely, from 0.5 to 45 days. Figure \ref{figHdeltaP} displays the behavior of $H$ versus $\Delta P/P$, from which one observes a strong trend of increasing $\Delta P/P$ towards longer $H$-index, paralleling the background found by different studies. This outcome is in agreement with the results found by \cite{reinhold}, where the relative differential rotation shear increases with longer rotation periods, as well as previous observations shown by \cite{barnes2005} and theoretical approaches \citep[e.g.,][]{kuker}. Finally, the comparison of the distribution of $H$ for active stars with one rotation period identified and active stars with two rotation periods identified using the Kolmogorov-Smirnov test, as shown above, reveals that a correlation between $\Delta P/P$ and $H$ can be claimed.

\section{Final remarks}
We have analyzed a homogeneous set of 8 332 active \textit{Kepler} stars presented in \cite{reinhold} and \cite{reinhold2015}. We calculated a new photometric activity index defined as the $H$-index for all the stars. To do so, we used two statistical methods denoted by Rescaled range ($R/S$) analysis and the MFDMA algorithm with the time series prepared by the PDC-msMAP pipeline. Special care was taken to remove all the pulsating stars in our sample. In general, we showed that the stars have a wide range of values of the $H$-index, indicating a variety of behaviors in the magnetic activity of the stars studied here.

Our final sample was divided by two different rotational behaviors into stars with differential rotation traces and those without differential rotation ones. By using from K-S test, we showed that the distributions of $H$ of stars with or without differential rotation does not from a same distribution. As an important result, the $H$- index is an able parameter for distinguishing the different signs of stellar rotation that can exist between the stars with and without differential rotation and consequently, a correlation between $\Delta P/P$ and $H$ can be claimed.

We found important differences in the rotation-$H$ relationship between the stars with and without differential rotation traces. These differences highlight the relevance of the variability range $R_{var}$ for interpreting the level of magnetic activity along the diagonal line described in Figures \ref{figHFractal} to \ref{figHr}. For stars without differential rotation traces, there is a clear division into two regimes of $R_{var}$ that are indicated by the sizes of the empty circles. In contrast, the stars with defined differential rotation showed an evident line, which we defined as a ``line of high activity''. This result shows that following this line, there is no distinction between slow and fast rotators. On the other hand, this distinction is clearer in the case of stars with rigid rotation. This result suggests that to investigate magnetic activity from the period of photometric rotational modulation, it is necessary to understand the dynamics of the long- and short-term persistence indicated by the Hurst exponent. However, detecting changes in the time series due to differential rotation is very difficult. Our methods have shown that the Hurst exponent is a promising index for estimating photometric magnetic activity. It corresponds to the first index for investigating the behavior of stellar rotation, which considers the dynamics of long- and short-term fluctuations.

We conclude that an analysis that incorporates the studied methods can add diagnostic power to contemporary analytic methods of time series analysis for studying the signatures of rigid bodies and those with differential rotations. We suggest the multifractal analysis as an alternative way that can help us to identify the source of differential rotation in active stars. We also suggest that the rotation--differential rotation relationship for the stars that are studied here is linked to the Hurst exponent due its strong correlation with rotation period \citep{defreitas2013}.

In summary, our suggestion that the dynamics of starspots for time series with and without differential rotation are distinct is an impactful result. In addition, our approach also suggests that the differential rotation signature as well as the rigid body one are explicitly governed by local fluctuations with smaller magnitudes, identified by a long-right tail of the multifractal spectrum inferred from the behavior of the degree of asymmetry ($A>1$ for most of the stars, as shown in Figure \ref{figA}). In general terms,  $A>1$ implies that the time series presents a strong rotational signature, modulated by the presence of spots, whereas the few stars found with $A<1$ shows a low signal to noise ratio. In the same line of reasoning, we identify the overall trend whereby differential rotation, which is represented by the parameter $\Delta P/P$, is correlated to the $H$-index segregated by the variability range $R_{var}$. This behavior agrees with other results found in the literature.

Finally, the results shown in the present work are not the final word on analyzing stellar rotation as a (multi)fractal process. Indeed, there is an outstanding question regarding the multifractal behaviors present in \textit{Kepler} time series that could motivate further research.

\acknowledgments
DBdeF acknowledges financial support 
from the Brazilian agency CNPq-PQ2 (grant No. 306007/2015-0). Research activities of STELLAR TEAM of Federal University of Cear\'a are supported by continuous grants from the Brazilian agency CNPq. JRM acknowledges CNPq, CAPES and FAPERN agencies for financial support. This paper includes data collected by the \textit{Kepler} mission. Funding for the \textit{Kepler} mission is provided by the NASA Science Mission directorate. All data presented in this paper were obtained from the Mikulski Archive for Space Telescopes
(MAST).

\end{document}